\documentclass[conference]{IEEEtran}
\IEEEoverridecommandlockouts
\pdfoutput=1
\usepackage{cite}
\usepackage{caption}
\usepackage{amsmath,amssymb,amsfonts}
\usepackage{bm}
\usepackage{steinmetz}
\usepackage{diagbox}
\usepackage{algorithmic,algorithm}
\usepackage{graphicx}
\usepackage{booktabs}
\usepackage{subfigure}
\usepackage{textcomp}
\usepackage{amsmath}
\usepackage{amsthm}
\usepackage{xcolor}
\usepackage[english]{babel}

\newtheorem{thm}{Theorem}

\newtheorem{assumption}{Assumption}

\def\BibTeX{{\rm B\kern-.05em{\sc i\kern-.025em b}\kern-.08em
		T\kern-.1667em\lower.7ex\hbox{E}\kern-.125emX
		
}}

\usepackage[top=0.69in, left=0.7in, bottom=1.03in, right=0.7in]{geometry}
\begin{document}
	\captionsetup[figure]{labelformat={default},labelsep=period,name={Fig.}}
	
	\title{STAR-RIS Assisted Over-the-Air Vertical Federated Learning in Multi-Cell Wireless Networks\vspace{-0.2cm}}
	
	\author{
		\IEEEauthorblockN{$\text{Xiangyu Zeng}^{\ast \dagger \ddagger}$, $\text{Yijie Mao}^{\ast}$, and $\text{Yuanming Shi}^{\ast}$
		}
		\IEEEauthorblockA{
			$^{\ast}$School of Information Science and Technology, ShanghaiTech University, Shanghai 201210, China \\ 
			$^{\dagger}$Shanghai Institute of Microsystem and Information Technology, Chinese Academy of Sciences, China\\
			$^{\ddagger}$University of Chinese Academy of Sciences, Beijing 100049, China\\
			E-mail: \{zengxy,  maoyj, shiym\}@shanghaitech.edu.cn }\vspace{-0.9cm}}
	\maketitle
	
	\begin{abstract}
		Vertical federated learning (FL) is a critical enabler for distributed artificial intelligence services in the emerging 6G era, as it allows for secure and efficient collaboration of machine learning among a wide range of Internet of Things devices. However, current studies of wireless FL typically consider a single task in a single-cell wireless network, ignoring the impact of inter-cell interference on learning performance. In this paper, we investigate a simultaneous transmitting and reflecting reconfigurable intelligent surface (STAR-RIS) assisted over-the-air computation based vertical FL system in multi-cell networks, in which a STAR-RIS is deployed at the cell edge to facilitate the completion of different FL tasks in different cells. We establish the convergence of the proposed system through theoretical analysis and introduce the Pareto boundary of the optimality gaps to characterize the trade-off among cells. Based on the analysis, we then jointly design the transmit and receive beamforming as well as the STAR-RIS transmission and reflection coefficient matrices to minimize the sum of the gaps of all cells. To solve the non-convex resource allocation problem, we introduce a successive convex approximation based algorithm. Numerical experiments demonstrate that compared with conventional approaches, the proposed STAR-RIS assisted vertical FL model and the cooperative resource allocation algorithm achieve much lower  mean-squared error for both uplink and downlink transmission in multi-cell wireless networks, resulting in improved learning performance for vertical FL.
	\end{abstract}
	
	\section{Introduction}
	Federated learning (FL) is a machine learning (ML) approach that enables multiple parties to collaboratively train a learning model without revealing their individual data. This is beneficial in a variety of fields where data privacy is a concern, as FL allows parties to maintain control over their own data while still benefiting from the combined knowledge of all parties. In modern wireless Internet of Things (IoT) networks, data is often collected from various types of devices \cite{9606720}. To facilitate data analysis in such settings, vertical FL, a variation of FL that is designed to address the challenges of training machine learning models on vertically partitioned data silos, is commonly adopted \cite{8808168,shi2020communication,wang2020wireless,yang2020federated,yang2022differentially}.
	
	One major issue that prevents the implementation of (vertical) FL in real-world application is the  communication latency. To address this issue, over-the-air computation (AirComp) has been proposed to facilitate fast wireless data aggregation. By utilizing the superposition property of wireless multiple access channels (MAC) to concurrently transmit and aggregate local updates, AirComp significantly reduces communication latency compared to orthogonal transmission. Previous research has explored the use of AirComp in FL, such as the joint design of device selection and beamforming for fast global model aggregation in \cite{yang2020federated}, and the development of a broadband analog aggregation scheme for low latency FL with linear growth of latency reduction ratio in \cite{zhu2019broadband}.
	
	On the other hand, the coexistence of multiple FL tasks in multi-cell networks has yet to be fully explored. Though the authors in \cite{xu2021bandwidth} have studied the bandwidth allocation for multiple FL tasks, the system model is limited to a single-cell network and the impact of inter-cell interference on FL performance remains unplumbed. It has been well investigated that reconfigurable intelligent surface (RIS), a metasurface composed of reconfigurable passive elements, can modify the propagation environment of wireless signal and  reduce multi-cell interference \cite{luo2021reconfigurable}. However, conventional RISs are reflecting only with limited wireless coverage \cite{huang2020holographic}. The recently introduced simultaneous transmitting and reflecting RIS (STAR-RIS), which allows the source and destination to be located at either side of the metasurface, has been recognized as a promising strategy to enhance the coverage of each cell and further reduce inter-cell interference \cite{liu2021star}. STAR-RIS is therefore a promising technique to facilitate FL in multi-cell networks. To the best of our knowledge, STAR-RIS assisted vertical FL has not been studied yet.
	
	In this paper, inspired by the benefits of AirComp for global aggregation \cite{9814484} and the merits of STAR-RIS in multi-cell networks, we fill the research gap and propose a STAR-RIS assisted AirComp-based vertical FL in multi-cell networks, where a STAR-RIS is deployed at the cell edge to assist each cell in completing different FL tasks. Through theoretical analysis, we demonstrate the convergence of our proposed vertical FL process and introduce the Pareto boundary of the gap region to characterize the trade-off performance among multiple cells. This allows us to formulate an optimization problem with the aim of minimizing the sum of error-induced gaps for all cells using the proposed algorithm based on successive convex approximation (SCA). Numerical experiments confirm the validity of our theoretical analysis and show the superiority of our proposed approach.

	\section{System Model}
	\subsection{Learning Framework}
	Consider a STAR-RIS assisted multi-cell wireless network consisting of $M$ base stations (BS) with $N$ antennas, where BS $m\in \mathcal{M} = \{1,2,\ldots,M\}$ aims to train an ML model by coordinating $K_{m}$ single-antenna devices located in cell $m$. Specifically, device $k \in \mathcal{K}_m=\left\{\sum_{l=1}^{m-1} K_l+1, \sum_{l=1}^{m-1} K_l+2, \ldots, \sum_{l=1}^{m-1} K_l+\right.$ $\left.K_m\right\}$ is associated with BS $m$. And there is one STAR-RIS equipped with $Q$ passive reflecting/transmitting elements, deployed at the cell-edge of all cells to boost the signal strength of edge devices. Each cell is equipped with a vertically partitioned dataset, where different devices hold different features of the same samples. For simplicity, we assume that each cell has the same number of samples and that devices within each cell contain the same number of non-overlapping features. Let $\mathcal{D}_m=\{(\bm{x}_{m,1}^{i}, \cdots, \bm{x}_{m,K_m}^{i}), y^{i}_m\}_{i=1}^{L_m}$ denote the whole training dataset of $L_m$ samples in cell $m$, where $\bm{x}_{m,k}^{i}$ denotes the partial features of sample $i$ located at device $k$ in cell $m$, and $y^i_m$ denotes the corresponding label. In vertical FL, it is assumed that the BS holds all labels $y_m=\{y^{i}_m\}_{i=1}^{L_m}$, and device $k$ is only available to its own local feature set $\mathcal{D}_{m,k}=\{\boldsymbol{x}_{m,k}^{i}\}_{i=1}^{L_m}.$ And $\boldsymbol{\bm{x}}_m^{i}=[(\boldsymbol{\bm{x}}_{m,1}^{i})^{\sf T}, \cdots,(\boldsymbol{\bm{x}}_{m,K_m}^{i})^{\sf T}]^{\sf T} $ denotes the overall feature vector of sample $i$. 
	
	The goal of vertical FL in cell $m$ is to collaboratively learn a global model $\bm{w}_m$ (concatenated vector of $\bm{w}_k$ for $k\in \mathcal{K}_m$) that maps an input  to the corresponding prediction through a continuously differentiable function $\sigma(\cdot)$. Since features of one sample are distributed at different devices, we assume that device $k$ maps the local feature $\boldsymbol{x}_{k}$ to local prediction result $g_{k}(\boldsymbol{w}_{k} ; \boldsymbol{x}_{k})$. This paper considers a linear form for the local prediction function, i.e., $g_{k}(\boldsymbol{w}_{k} ; \boldsymbol{x}_{k})=\boldsymbol{w}_{k}^{\sf T} \boldsymbol{x}_{k}.$ By aggregating local prediction results, the final prediction in cell $m$ can be obtained by $
	\sigma(\boldsymbol{w}_m ; \boldsymbol{x}_m)=\sigma(\sum_{k\in \mathcal{K}_m} g_{k}(\boldsymbol{w}_{k} ; \boldsymbol{x}_{k})) \label{prediction} =\sigma(\bm{w}_m^{\sf T}\bm{x}_m).$ 
	In order to learn the global model $\bm{w}_m$ in cell $m$, we propose to minimize the loss function as
	\vspace{-0.1cm}
	\begin{equation}
		\min _{\boldsymbol{w}_m } F(\boldsymbol{w}_m)=\frac{1}{L_m} \sum_{i=1}^{L_m} f\left(\sigma(\boldsymbol{w}_m^{\sf T} \boldsymbol{x}_m^{i}) ; y_m^{i}\right),\label{loss}
		\vspace{-0.1cm}
	\end{equation}
	where $f(\cdot)$ is the sample-wise loss function.
	
	In our multi-cell system, each cell performs a unique FL task using the full batch gradient descent (GD) approach, which is described in the following subsection. We assume universal frequency reuse, meaning that all cells share the same frequency channel, leading to inter-cell interference.
	\vspace{-0.1cm}
	\subsection{GD Algorithm for Vertical FL}
	In this subsection, we introduce the framework of GD algorithm for vertical FL. For brevity, the subscript of cell $m$ is omitted for $L_m,\mathbf{w}_m,\mathbf{x}_m, y_m$. The GD algorithm specified in this subsection is applied for all cells. Let $\nabla F(\boldsymbol{w})$ denote the gradient of $F$ respect to $\boldsymbol{w}$, which is calculated as
	\vspace{-0.1cm}
	\begin{equation}
		\nabla F(\boldsymbol{w})=\frac{1}{L} \sum_{i=1}^{L} \nabla f(\sigma(\boldsymbol{w}^{\sf T} \boldsymbol{x}^{i}) ; y^{i}),
	\end{equation}
	where $\nabla f(\sigma(\boldsymbol{w}^{\sf T} \boldsymbol{x}^{i}) ; y^{i})$ denote the gradient of $f(\sigma(\boldsymbol{w}^{\sf T} \boldsymbol{x}^{i}) ; y^{i})$  respect to $\boldsymbol{w}$. 
	Based on the chain rule, the gradient of $f$ is rewritten as
	\begin{equation}
		\nabla f(\sigma(\boldsymbol{w}^{\sf T} \boldsymbol{x}^{i}) ; y^{i}) =G(\boldsymbol{w}^{\sf T} \boldsymbol{x}^{i} ; y^{i}) \boldsymbol{x}^{i},
	\end{equation}
	where $G(\boldsymbol{w}^{\sf T} \boldsymbol{x}^{i} ; y^{i})=\partial f(\sigma(\boldsymbol{w}^{\sf T} \boldsymbol{x}^{i}) ; y^{i}) / \partial \boldsymbol{w}^{\sf T} \boldsymbol{x}^{i}$ is an auxiliary function. Hence, $\nabla F(\boldsymbol{w})$ can be rewritten as
	\begin{equation}
		\begin{aligned}
			\nabla F(\boldsymbol{w}) &=\frac{1}{L} \sum_{i=1}^{L} G(\boldsymbol{w}^{\sf T} \boldsymbol{x}^{i} ; y^{i})\bm{x}^i.
		\end{aligned}
	\end{equation}
	Recall that the BS holds all labels $y$, so $G(\boldsymbol{w}^{\sf T} \boldsymbol{x}^{i} ; y^{i})$ can be calculated at the BS only if the BS can access the aggregation of local predictions $\{\boldsymbol{w}^{\sf T} \boldsymbol{x}^{i}\}_{i=1}^{L}$. Specifically, at the $t$-th communication round, the BS and the edge devices in each cell perform the following three procedures:
	
	\textbf{Broadcasting:} The BS computes $\{G((\bm{w}^{(t)})^{\sf T}\bm{x}^i;y^i)\}_{i=1}^L$ and broadcasts the result back to its corresponding devices.
	
	\textbf{Local model update:} After broadcasting, device $k$ computes the partial gradient $\nabla_{k} F(\boldsymbol{w}_k)$ with local data $\mathcal{D}_k$, given as
	\begin{equation}
		\nabla_{k} F(\boldsymbol{w}_k)=\frac{1}{L} \sum_{i=1}^{L} G(\boldsymbol{w}^{\sf T} \boldsymbol{x}^{i} ; y^{i})  \boldsymbol{x}_{k}^{i} \label{gradient_k} .
	\end{equation}
	Each device can thus update its local model by taking a step of GD with learning rate $\mu^{(t)}$ as
	\begin{equation}
		\boldsymbol{w}_{k}^{(t+1)}=\boldsymbol{w}_{k}^{(t)}-\mu^{(t)} \nabla_{k} F(\boldsymbol{w}_{k}^{(t)}),\label{gd}
	\end{equation}
	where $\bm{w}_k^{(t)}$ is the local model of device $k$ at the $t$-th round.
	
	\textbf{Local prediction and global aggregation:} device $k$ computes the local prediction results $\{(\boldsymbol{w}_k^{(t+1)})^{\sf T} \boldsymbol{x}^{i}_k\}_{i=1}^{L}$ and sends to the BS. And BS aggregates them to get final prediction result $\{(\boldsymbol{w}^{(t+1)})^{\sf T} \boldsymbol{x}^{i}\}_{i=1}^{L}$.
	
	Since the BS only needs the aggregation of local prediction results, i.e., neither local features nor local models need be uploaded to the BS, which significantly enhances privacy protection. In addition, the communication efficiency is improved since the local prediction result is usually low-dimensional.
	
	\subsection{Communication Model} 
	In this subsection, the proposed communication model is delineated with a special focus on the STAR-RIS assisted uplink and downlink transmission models.
	\subsubsection{STAR-RIS}
	The STAR-RIS  is a type of RIS that can produce omnidirectional radiation by implementing equivalent electric and magnetic currents in its hardware. It has three protocols for use in wireless networks: energy splitting, mode switching, and time switching. In this article, we focus on the mode-switching protocol, in which each element of the STAR-RIS can operate in either the reflection mode (R mode) or the transmission mode (T mode). Such on-off type of operating protocol is simpler to implement compared to the energy splitting protocol. Specifically, one group consists of $Q^t$ elements operating in the T mode, while the other group contains $Q^r$ elements operating in the R mode, where $Q^t+Q^r=Q$. Accordingly, the STAR-RIS transmission-coefficient and reflection-coefficient matrices are given by $\bm\Theta_t=$ $\operatorname{diag}\left(\sqrt{\beta_1^t} e^{j \theta_1^t}, \sqrt{\beta_2^t} e^{j \theta_2^t}, \ldots, \sqrt{\beta_Q^t} e^{j \theta_Q^t}\right)$ and $\bm\Theta_r=\operatorname{diag}\left(\sqrt{\beta_1^r} e^{j \theta_1^r}, \sqrt{\beta_2^r} e^{j \theta_2^r}, \ldots, \sqrt{\beta_Q^r} e^{j \theta_Q^r}\right)$, respectively, where $\beta_q^t, \beta_q^r \in\{0,1\}, \beta_q^t+\beta_q^r=1$, and $\theta_q^t, \theta_q^r \in[0,2 \pi), \forall q \in \{1,2,\ldots,Q\}$. The $M$ cells can be divided into two groups  $\mathcal{M}_r$ and $\mathcal{M}_t$. Specifically, cell $m$ is in the reflection dimension with $m \in \mathcal{M}_r$ and in the transmission dimension with $m\in \mathcal{M}_t$.
	
	Let $\boldsymbol{h}_{m, k} \in \mathbb{C}^N, \boldsymbol{h}_{k}^r \in \mathbb{C}^{Q}$ and $\boldsymbol{G}_m\in \mathbb{C}^{Q \times N}$ denote the equivalent channels from edge device $k$ to BS $m$, from edge device $k$ to the STAR-RIS, and from the STAR-RIS to BS $m$, respectively. The combined channel from the $k$-th edge device to the BS $m$ via the STAR-RIS can be written as
	$$\bar{\bm h}_{m,k}=\left\{
	\begin{array}{rcl}
		\bm{h}_{m,k} + \bm{G}_m^{\sf H}\bm\Theta_t\bm {h}_{k}^r, \forall  m \in \mathcal{M}_t,  \\ 
		\bm{h}_{m,k} + \bm{G}_m^{\sf H}\bm\Theta_r\bm {h}_{k}^r, \forall m \in \mathcal{M}_r.
	\end{array}
	\right. $$
	Note that the uplink and downlink STAR-RIS matrices can be separatively designed. For simplify, we write $\bm\Theta_t$ and $\bm\Theta_r$ for uplink and downlink transmission in terms of $\bm\Theta^{\text{ul}}$ and $\bm\Theta^{\text{dl}}$, respectively.
	\subsubsection{Uplink transmission}
	In the uplink transmission, we assume the devices communicate with the BS via AirComp, which has a wide range of FL applications.
	
	Specifically, we denote $\bm{s}_k = [s_k^1, s_k^2, \cdots, s_k^L]^{\sf T}\in \mathbb{C}^L$ as the local prediction results at device $k$, where the local prediction result of the $i$-th sample $s_k^{i} = \bm{w}_k^{\sf T}\bm{x}_k^i.$ At each time slot $i\in\{1,2,\cdots, L_m\}$, each device in cell $m$ sends the corresponding prediction result of the $i$-th sample to BS $m$. And we assume that $\bm{s}_k$ is normalized with zero mean and unit variance \cite{liu2021reconfigurable}. We denote $g_m(i) = \sum_{k\in \mathcal{K}_m} s_k^i$ as the target function to be estimated through AirComp at the $i$-th time slot. To simplify the notation, we omit the time index by writing $g(i)$ and $s_k^i$ as $g$ and $s_k^{\text{ul}}$, respectively. And we assume that the signals transmitted by all devices are synchronized at the BS. Then the received signal at BS $m$ is given by
	\begin{equation}
		\begin{aligned}
			\bm{y}_m^{\text{ul}}=& \sum_{k}\bar{\bm{h}}_{m,k} b_{k}s_k^{\text{ul}} + \bm{n}_m^{\text{ul}},
		\end{aligned}
	\end{equation}
	where $b_{k} \in \mathbb{C}$ is the transmit scalar at device $k$,  and $\bm{n}_m^{\text{ul}}$ is the additive white Gaussian noise with zero mean and variance $(\sigma^{\text{ul}})^2$ at BS $m$. The transmit power constraint at device $k$ is $\mathbb{E}(|b_ks_k^{\text{ul}}|^2) = |b_k|^2 \leq P^{\text{ul}},$ where $ P^{\text{ul}} > 0$ is the maximum transmit power. The scaled signal received at BS $m$ is
	\begin{equation}
		\bar{g}_m =\frac{1}{\sqrt{\eta_m}} \boldsymbol{r}_m^{\sf H} \boldsymbol{y}_m^{\text{ul}}=\frac{1}{\sqrt{\eta_m}} \boldsymbol{r}_m^{\sf H} \sum_{k\in \mathcal{K}} \bar{\boldsymbol{h}}_{m,k} b_{k} s_{k}^{\text{ul}}+\frac{\boldsymbol{r}_m^{\sf H} \boldsymbol{n}_m^{\text{ul}}}{\sqrt{\eta_m}}, 
	\end{equation}
	where $\bm{r}_m \in \mathbb{C}^{N}$ is the receive beamforming vector and $\eta_m$ is a normalizing factor for cell $m$. To compensate for the phase distortion introduced by complex channel responses, the transmit scalar at device $k$ in cell $m$ is set to $b_k = \sqrt{\eta_m}\frac{(\bm{r}_m^{\sf H}\bar{\bm{h}}_{m,k})^{\sf H}}{|\bm{r}_m^{\sf H}\bar{\bm{h}}_{m,k}|^2}, \forall k\in \mathcal{K}_m,$ and $\eta_m$ can be expressed as $\eta_m=P^{\text{ul}} \min _{k \in \mathcal{K}_m}|\boldsymbol{r}_m^{\sf H} \bar{\boldsymbol{h}}_{m,k}|^{2}.$ Then the estimated function at BS for cell $m$ is given as
	\begin{equation}
		\resizebox{0.88\hsize}{!}{$
			\begin{aligned}
				\hat{g}_m &=\Re \{\bar{g}_m\} \\
				&=\Re \{ g_m + \underbrace{\frac{1}{\sqrt{\eta_m}}\bm{r}_m^{\sf H}\sum_{l\neq m}\sum_{j\in \mathcal{K}_l}\bar{\boldsymbol{h}}_{m,j} b_{j} s_{j}^{\text{ul}}  +\frac{\boldsymbol{r}_m^{\sf H} \boldsymbol{n}_m^{\text{ul}}}{\sqrt{\eta_m}} }_{e_m^{\text{ul}}}\} \\
				&= g_m + \Re\{e_m^{\text{ul}}\}
			\end{aligned}
			$}
	\end{equation}
	
	\subsubsection{Downlink transmission}
	After obtaining the estimate $\hat{g}_m$ in the cell $m$, BS $m$ computes $G(\hat{g}_m;y)$ with noisy aggregation $\hat{g}_m$, and then broadcasts the result to the associated devices in $\mathcal{K}_m$. And we write $G(\hat{g}_m;y)$ in terms of $G_m$ for simplify. Without loss of generality, we assume that the transmitted signal follows the standard Gaussian distribution, i.e., $G_m \sim \mathcal{CN}(0,1)$. The received signal at device $k$ is
	\begin{equation}
		y_k^{\text{dl}} = \sum_m\bar{\bm{h}}_{m,k}^{\sf H}\bm{t}_m G_m + n_k^{\text{dl}},
	\end{equation}
	where $\bm{t}_m$ denotes the transmit beamforming vector at BS $m$, and $n^{\text{dl}}_k \sim \mathcal{C} \mathcal{N}\left(0, (\sigma^{\text{dl}})^2 \right)$ is the additive white Gaussian noise with zero mean and variance $(\sigma^{\text{dl}})^2$ at device $k$.  The maximum transmit power at BS $m$ is $P^{\text{dl}}$, i.e., $\mathbb{E}(\|G_m\bm{t}_m\|^2) = \|\bm{t}_m\|^2 \leq P^{\text{dl}}$.
	
	To compensate for the phase distortion introduced by complex channel responses, the receive scalar at device $k$ in cell $m$ is set to $r_{k}=\frac{(\bar{\bm{h}}_{m,k}^{\sf H}\bm{t}_m)^{\sf H}}{\left|\bar{\bm{h}}_{m,k}^{\sf H}\bm{t}_m\right|^2} $. The estimated $G_m$ at device $k$ is given as
	\begin{equation}
		\resizebox{0.88\hsize}{!}{$
			\begin{aligned}
				\hat{G}_{m,k} &=\Re \{r_k y_k^{dl}\} \\
				&=\Re \{ G_m + \underbrace{\frac{(\bar{\bm{h}}_{m,k}^{\sf H}\bm{t}_m)^{\sf H}}{\left|\bar{\bm{h}}_{m,k}^{\sf H}\bm{t}_m\right|^2} \left(\sum_{l \neq m}  \bar{\bm{h}}_{l,k}^{\sf H} \bm{t
					}_l G_l + n_k^{\text{dl}}\right)}_{\bar{e}_k^{\text{dl}}}\} \\
				&= G_m + \Re\{\bar{e}_k^{\text{dl}}\}.
			\end{aligned}
			$}%
	\end{equation}
	Note that the uplink noise is embedded in function $G_m$.  In order to directly describe the effective noise, we expand $G_m$ to its first-order Taylor expansion as follows
	\begin{equation}
		\label{taylor}
		\begin{aligned}
			&\hat{G}_{m,k} \\
			&= G_m + \Re\{\bar{e}_k^{\text{dl}}\} \\
			&= G(g_m+\Re\{e_m^{\text{ul}}\};y) + \Re\{\bar{e}_k^{\text{dl}}\} \\
			&= G(g_m;y) + G^{'}(g_m ; y)\Re\{e_m^{\text{ul}}\} + \mathcal{O}(|\Re\{e_m^{\text{ul}}\}|^2) + \Re\{\bar{e}_k^{\text{dl}}\}  \\
			&\approx G(g_m;y) + \underbrace{G^{'}(g_m ; y)\Re\{e_m^{\text{ul}}\}+ \Re\{\bar{e}_k^{\text{dl}}\}}_{e_k^{\text{dl}}},
		\end{aligned}
	\end{equation}
	where $G^{\prime}(\cdot)$ is the first derivative of $G(\cdot)$. Assume that the noise amplitude is small, the term $\mathcal{O}(|\Re\{e_m^{\text{ul}}\}|^2)$ is neglected, which implies the last approximation in \eqref{taylor}.
	
	\section{Convergence Analysis and Problem Formulation}
	\subsection{Convergence Analysis}
	In previous work \cite{wang2021federated, li2019convergence,9814484}, the convergence analysis of the AirComp-based vertical FL process in each cell has been established under the following assumptions. 
	\begin{assumption}[$\alpha$-strongly convexity] \label{asm1}
		The function $F(\cdot)$ is assumed to be $\alpha$-strongly convex on $\mathbb{R}^{d}$ with constant $\alpha$, namely, for all $\boldsymbol{x}, \boldsymbol{y} \in \mathbb{R}^{d}$, we have
		\begin{equation}
			F(\boldsymbol{y}) \geq F(\boldsymbol{x})+\nabla F(\boldsymbol{x})^{\sf T}(\boldsymbol{y}-\boldsymbol{x})+\frac{\alpha}{2}\|\boldsymbol{y}-\boldsymbol{x}\|_2^{2}.  \notag
		\end{equation}
	\end{assumption}
	\begin{assumption}[$\beta$-smoothness] \label{asm2}
		The function $F(\cdot)$ is assumed to be $\beta$-smooth on $\mathbb{R}^{d}$ with constant $\beta$, namely, for all $\boldsymbol{x}, \boldsymbol{y} \in \mathbb{R}^{d}$, we have
		\begin{equation}
			F(\boldsymbol{y}) \leq F(\boldsymbol{x})+\nabla F(\boldsymbol{x})^{\sf T}(\boldsymbol{y}-\boldsymbol{x})+\frac{\beta}{2}\|\boldsymbol{y}-\boldsymbol{x}\|_2^{2}.  \notag
		\end{equation}
	\end{assumption}
	\begin{thm}[Convergence of vertical FL process]
		\label{thm1}
		Suppose that \textbf{Assumption} \ref{asm1} and \ref{asm2} hold, setting the learning rate to be $0<\mu^{(t)}\leq \frac{1}{\beta}$, then the expected optimality gap after T communication rounds is upper bounded by
		\begin{equation}
			\resizebox{1.02\hsize}{!}{$
				\begin{aligned}
					&\mathbb{E}\left[F(\bm{w}_m^{(T)}) - F(\bm{w}_m^*)\right] \leq \rho^T \mathbb{E}\left[F(\bm{w}_m^{(0)})-F(\bm{w}_m^*)\right] \\
					&+\frac{1}{2 \beta L^2} \sum_{t=0}^{T-1} \rho^{T-t-1} \sum_{k\in \mathcal{K}_m}\left(\Phi_{1,k}\mathbb{E}[|\Re\{e_{m}^{\text{ul}}\}|^2] + \Phi_{2,k}\mathbb{E}[|\Re\{\bar{e}_{k}^{\text{dl}}\}|^2]  \right),
				\end{aligned}
				$}
		\end{equation}
		where $\rho=1-\alpha / \beta$, $\Phi_{1,k} = \sum_{i=1}^L\|(G^{i}_{m,k})^{'}\bm{x}_k^i\|_2^2$ and $\Phi_{2,k} = \sum_{i=1}^L\|\bm{x}_k^i\|_2^2$.
		\begin{proof}
			Please refer to previous work \cite{9814484}.
		\end{proof}
	\end{thm}
	
	\subsection{Problem Formulation}
	According to $\textbf{Theorem \ref{thm1}}$, the convergence optimality gap is largely determined by the mean-squared-error (MSE) of both $g_m$ and $G_m$. However, solely optimizing MSE for each cell through AirComp may result in significant inter-cell interference in the considered multi-cell wireless networks, which can negatively impact the learning performance of other cells. As such, it is necessary to carefully balance the learning performance among various FL tasks in multiple cells through a cooperative design.
	
	We begin by identifying the gap region $\mathcal{G}$, to be the set of tuples $\left(\Delta_{1}, \Delta_{2}, \ldots, \Delta_{M}\right)$, which represents the instantaneous errors that cause gaps in all cells, and can be achieved simultaneously under specific downlink and uplink transmission power constraints. The gap region $\mathcal{G}$ can be 
	represented as
	\begin{equation}
		\mathcal{G} = \bigcup \{(\Delta_{1}, \Delta_{2}, \ldots, \Delta_{M}) | \Delta_m \geq \text{Gap}_m, \forall m\in \mathcal{M}  \},
	\end{equation}
	where
	\begin{equation}
		\text{Gap}_m = \sum_{k\in \mathcal{K}_m}\left(\Phi_{1,k}\mathbb{E}[|\Re\{e_{m}^{\text{ul}}\}|^2] + \Phi_{2,k}\mathbb{E}[|\Re\{\bar{e}_{k}^{\text{dl}}\}|^2]  \right),
	\end{equation}
	\begin{equation}
		\begin{aligned}
			&\mathbb{E}[|\Re\{e_{m}^{\text{ul}}\}|^2] = \sum_{l\neq m,j\in \mathcal{K}_l} \frac{\eta_l|\bm{r}_m^{\sf H}\bar{\bm{h}}_{m,j} |^2}{\eta_m |\bm{r}_l^{\sf H}\bar{\bm{h}}_{l,j}|^2} + \frac{\|\bm{r}_m \|^2\sigma_{\text{ul}}^2}{\eta_m}, \\
			&\mathbb{E}[|\Re\{\bar{e}_{k}^{\text{dl}}\}|^2]= \frac{\sum_{l \neq m}   |\bar{\bm{h}}_{l,k}^{\sf H} \bm{t}_l  |^{2}
				+ (\sigma^{\text{dl}})^2}{\left|\bar{\bm{h}}_{m,k}^{\sf H}\bm{t}_m\right|^2}.
		\end{aligned}
	\end{equation}
	
	As previously stated, in order to decrease the error-induced gap in one cell, the gaps of other cells maybe increased. In light of this, our objective is to find a suitable solution that allows us to achieve the Pareto boundary of the gap region $\mathcal{G}$, so as to balance the performance of learning among multiple cells. In this context, the Pareto optimality of a tuple is described as follows\cite{jorswieck2008complete}.
	
	Here, we leverage the profiling technique \cite{cao2020cooperative} to characterize the Pareto boundary by coordinating all BSs to minimize the sum of Gap of all cells. Specifically, let $\kappa=\left[\kappa_{1}, \kappa_{2}, \ldots, \kappa_{M}\right]$ denote a given profiling vector, which satisfies $\kappa_{m} \geq 0, \forall m \in \mathcal{M}$, and $\sum_{m \in \mathcal{M}} \kappa_{m}=$ 1. The gap tuple on Pareto boundary can be obtained by solving the following problem
	\begin{subequations}
		\label{origin}
		\begin{align}
			\underset{\zeta,\{\bm{r}_{m}\},\{\bm{t}_m\}, \bm{\Theta}_t, \bm{\Theta}_r }{\operatorname{minimize}} \quad& \zeta \\
			\text { s.t. }\quad& \text{Gap}_m \leq \kappa_{m} \zeta, \forall m\in \mathcal{M}\\
			&\zeta \geq 0,
		\end{align}
	\end{subequations}
	where $\zeta$ denotes the sum of the gaps of all cells. Thus, the gap tuple can be represented as $\left(\Delta_{1}, \Delta_{2}, \ldots, \Delta_{M}\right)=\left(\kappa_{1} \zeta, \kappa_{2} \zeta, \ldots, \kappa_{M} \zeta\right)$, where a smaller value of $\kappa_{m}$ implies a more stringent requirement for the gap of cell $m$.
	
	Denote $\zeta = \zeta^{ul}+\zeta^{dl}$, where $\zeta^{ul}$ and $\zeta^{dl}$
	are used to quantify
	the sum of instantaneous error-induced gaps generated by
	uplink and downlink transmissions, respectively. Hence, we rewrite problem \eqref{origin} as 
	\begin{subequations}
		\label{ul+dl}
		\begin{align}
			\underset{\zeta^{ul},\zeta^{dl},\{\bm{r}_{m}\},\{\bm{t}_m\}, \bm{\Theta}_t, \bm{\Theta}_r}{\operatorname{minimize}} \quad& \zeta^{ul}+\zeta^{dl} \\
			\text { s.t. }\quad& \text{Gap}_m^{ul} \leq \kappa_{m} \zeta^{ul}, \forall m\in \mathcal{M}\\
			& \text{Gap}_m^{dl} \leq \kappa_{m} \zeta^{dl}, \forall m\in \mathcal{M}\\
			&\zeta^{ul} \geq 0\\
			&\zeta^{dl} \geq 0.
		\end{align}
	\end{subequations}
	The downlink and uplink transmissions can be decoupled in
	problem \eqref{ul+dl}, which allows us to separately optimize the
	downlink and uplink transmission resources.
	
	\section{optimization framework}
	In this section, we specify the optimization framework for solving the uplink and downlink optimization problems, respectively.
	\subsection{Uplink Optimization}
	For the uplink  aggregation, the optimization problem is
	\begin{subequations}
		\begin{align}
			\underset{\zeta^{ul},\{\bm{r}_{m}\},\bm{\Theta}^{ul}}{\operatorname{minimize}} \quad& \zeta^{ul} \\
			\text { s.t. }\quad& \sum_{l\neq m}\sum_{j\in \mathcal{K}_l}  \frac{\eta_l|\bm{r}_m^{\sf H}\bar{\bm{h}}_{m,j} |^2}{\eta_m |\bm{r}_l^{\sf H}\bar{\bm{h}}_{l,j}|^2} \\
			&+\frac{ \|\boldsymbol{r}_m\|^{2} (\sigma^{\text{ul}})^2}{\eta_m} \leq \kappa_{m} \zeta^{ul}, \forall m\in \mathcal{M}\\
			&\zeta^{ul} \geq 0 \label{zeta0}.
		\end{align}
	\end{subequations}
	By setting optimzing varibales $\bm{q}_{i}= \bm{r}_{i} /\sqrt{\eta_i}, \forall i \in \mathcal{M} $, the problem can be converted to
	\begin{subequations}
		\begin{align}
			\underset{\zeta^{ul},\{\bm{q}_{m}\},\bm{\Theta}^{ul}}{\operatorname{minimize}} \quad& \zeta^{ul} \\
			\text { s.t. }\quad& \sum_{l\neq m}\sum_{j\in \mathcal{K}_l}  \frac{|\bm{q}_m^{\sf H}\bar{\bm{h}}_{m,j} |^2}{ |\bm{q}_l^{\sf H}\bar{\bm{h}}_{l,j}|^2} \notag \\
			&+(\sigma^{\text{ul}})^2 \|\boldsymbol{q}_m^{\sf H}\|^{2}    \leq \kappa_{m} \zeta^{ul}, \forall m\in \mathcal{M}  \\
			&|\bm{q}_m^{\sf H}\bar{\bm{h}}_{m,k} |^2 \geq \frac{1}{P_{\text{ul}}}, \forall m,\forall k\in \mathcal{K}_m \label{cons_q} \\
			& \eqref{zeta0}.  \notag
		\end{align}
	\end{subequations}
	Then we let $\frac{|\bm{q}_m^{\sf H}\bar{\bm{h}}_{m,j} |^2}{ |\bm{q}_l^{\sf H}\bar{\bm{h}}_{l,j}|^2} \leq b_{l,j}$, the optimization problem relaxes to 
	\begin{subequations}\label{ul_pro}
		\begin{align}
			\underset{\zeta^{ul},\{\bm{q}_{m}, b\},\bm{\Theta}^{ul}}{\operatorname{minimize}} \quad& \zeta^{ul} \\
			\text { s.t. }\quad& \sum_{l\neq m}\sum_{j\in \mathcal{K}_l} b_{l,j}
			+ (\sigma^{\text{ul}})^2 \|\boldsymbol{q}_m^{\sf H}\|^{2}  \leq \kappa_{m} \zeta^{ul},  \forall m \label{cons_gap} \\
			&\frac{|\bm{q}_m^{\sf H}\bar{\bm{h}}_{m,j} |^2}{ |\bm{q}_l^{\sf H}\bar{\bm{h}}_{l,j}|^2} \leq b_{l,j}, \forall l,j  \label{cons_b} \\
			&  \eqref{zeta0},\eqref{cons_q} \notag .
		\end{align}
	\end{subequations} 
	However, constraint \eqref{cons_b} is still non-convex, then we use the SCA method to transform \eqref{cons_b} into a linear constraint which satisfies the property of convex. Let  $\boldsymbol{a}_{l,j}= [\Re(\bm{q}_l^{\sf H}\bar{\bm{h}}_{l,j}),\Im(\bm{q}_l^{\sf H}\bar{\bm{h}}_{l,j})]$, the corresponding approximated linear constraint is
	\begin{equation}
		\label{sca}
		\begin{split}
			\frac{|\bm{q}_m^{\sf H}\bar{\bm{h}}_{m,j} |^2}{b_{l,j}}  &\leq  \|\boldsymbol{a}_{l,j}\|^2   \\
			&\leq  	\|\boldsymbol{a}_{l,j}^{(t)}\|^2 + 2 (\boldsymbol{a}_{l,j}^{(t)})^{\sf T} (\boldsymbol{a}_{l,j}-\boldsymbol{a}_{l,j}^{(t)})
		\end{split}
	\end{equation}
	and
	\begin{equation}
		\|\boldsymbol{a}_{m,k}^{(t)}\|^2 + 2 (\boldsymbol{a}_{m,k}^{(t)})^{\sf T} (\boldsymbol{a}_{m,k}-\boldsymbol{a}_{m,k}^{(t)}) \geq \frac{1}{P_{\text{ul}}}. \label{cons_q_f}
	\end{equation}
	The origin problem \eqref{ul_pro} is then approximated as
	\begin{equation}\label{ul_pro_f}
		\begin{split}
			\underset{\zeta^{ul},\{\bm{q}_{m}, b, \bm{a}\},\bm{\Theta}^{ul}}{\operatorname{minimize}} \quad& \zeta^{ul} \\
			\text { s.t. }\quad&\boldsymbol{a}_{l,j}= [\Re(\bm{q}_l^{\sf H}\bar{\bm{h}}_{l,j}),\Im(\bm{q}_l^{\sf H}\bar{\bm{h}}_{l,j})], \forall l,j\\
			& \eqref{zeta0}, \eqref{cons_gap}, \eqref{sca},\eqref{cons_q_f}.
		\end{split}
	\end{equation} 
	And we can observe that the above problem turns out to be highly intractable due to the non-convexity of multiplication between variables $\bm{q}$ and $\bm\Theta^{ul}$. Hence, a classical alternative optimization algorithm  can be used to solve it.
	
	\subsection{Downlink Optimization
	}
	For the downlink dissemination,  the optimization problem can be written as
	\begin{subequations}
		\begin{align}
			\underset{\zeta^{dl},\{\bm{t}_{m}\},\bm\Theta^{dl}}{\operatorname{minimize}} \quad& \zeta^{dl} \\
			\text { s.t. }\quad&\sum_{k\in\mathcal{K}_m} \frac{\sum_{l \neq m}   |\bar{\bm{h}}_{l,k}^{\sf H} \bm{t}_l  |^{2}
				+ (\sigma^{\text{dl}})^2}{\left|\bar{\bm{h}}_{m,k}^{\sf H}\bm{t}_m\right|^2} \leq \kappa_{m} \zeta^{dl}, \forall m\\
			&\|\bm{t}_m \|^2 \leq P^{\text{dl}}  ,\label{cons_power} \\
			&\zeta^{dl} \geq 0 \label{zeta_dl}.
		\end{align}
	\end{subequations}
	By letting $ \frac{\sum_{l \neq m}   |\bar{\bm{h}}_{l,k}^{\sf H} \bm{t}_l  |^{2}
		+ (\sigma^{\text{dl}})^2}{\left|\bar{\bm{h}}_{m,k}^{\sf H}\bm{t}_m\right|^2} \leq d_{k}$, the optimization problem is relaxed to
	\begin{subequations}
		\begin{align}
			\underset{\zeta^{dl},\{\bm{t}_{m}\},\bm\Theta^{dl}}{\operatorname{minimize}} \quad& \zeta^{dl} \\
			\text { s.t. }\quad&\sum_{k\in\mathcal{K}_m} d_k  \leq \kappa_{m} \zeta^{dl}, \forall m\in \mathcal{M} \label{d}\\ 
			&\frac{\sum_{l \neq m}   |\bar{\bm{h}}_{l,k}^{\sf H} \bm{t}_l  |^{2}
				+ (\sigma^{\text{dl}})^2}{\left|\bar{\bm{h}}_{m,k}^{\sf H}\bm{t}_m\right|^2} \leq d_{k}, \label{cons_d} \\
			&\eqref{cons_power},\eqref{zeta_dl}. \notag
		\end{align}
	\end{subequations}
	Similar as the uplink optimization, we can still convert \eqref{cons_d} to linear constraints using the SCA method. By setting $\boldsymbol{c}_{m,k} =[\Re(\bar{\bm{h}}_{m,k}^{\sf H}\bm{t}_m),\Im(\bar{\bm{h}}_{m,k}^{\sf H}\bm{t}_m)]$, the relaxed problem is given as
	\begin{subequations}\label{dl_p}
		\begin{align}
			\underset{\zeta^{dl},\{\bm{t}_{m},\boldsymbol{c}\},\bm\Theta^{dl}}{\operatorname{minimize}} \quad& \zeta^{dl} \\
			\text { s.t. }
			& \frac{\sum_{l \neq m}   |\bar{\bm{h}}_{l,k}^{\sf H} \bm{t}_l  |^{2}
				+ (\sigma^{\text{dl}})^2}{d_k} \leq \notag \\
			&\|\boldsymbol{c}_{m,k}^{(t)}\|^2 + 2 (\boldsymbol{c}_{m,k}^{(t)})^{\sf T} (\boldsymbol{c}_{m,k}-\boldsymbol{c}_{m,k}^{(t)}), \forall m,k \\				
			& \boldsymbol{c}_{m,k} =[\Re(\bar{\bm{h}}_{m,k}^{\sf H}\bm{t}_m),\Im(\bar{\bm{h}}_{m,k}^{\sf H}\bm{t}_m)],\forall m,k \\
			& \eqref{cons_power}, \eqref{zeta_dl}, \eqref{d} \notag.
		\end{align}
	\end{subequations}
	The above problem \eqref{dl_p} can be solved in the same way as uplink optimization.
	
	\section{Simulation Results}
	In this section, we conduct extensive numerical experiments to evaluate the performance of the proposed SCA algorithm for the STAR-RIS assisted AirComp-based vertical FL system in multi-cell wireless network.
	
	We consider a STAR-RIS assisted two-cell wireless vertical FL network in a two-dimensional space, where the coordinates of the BSs are $(0\text{m},0\text{m})$ and $(40\text{m},0\text{m})$, the STAR-RIS is deployed at the edge of two cells, i.e., $(20\text{m},0\text{m})$.  And the devices in each cell are uniformly located within a circular region centered at their corresponding BS with radius $20$ meters.  All channel coefficients are modeled as
	\begin{equation}
		\bm{h}=\rho^{-\alpha / 2}\left(\sqrt{\frac{\beta}{1+\beta}} \bm{h}_{\mathrm{LoS}}+\sqrt{\frac{1}{1+\beta}} \bm h_{\mathrm{NLoS}}\right)
	\end{equation}
	and vary independently over different rounds, where $\rho$ denotes the distance between the transmitter and the receiver, $\alpha=2.5$ denotes the pathloss exponent, $\beta=5\mathrm{~dB}$ represents the Rician factor, $\bm h_{\text {LoS }}$ denotes the line-of-sight $(\mathrm{LoS})$ component, and $\bm h_{\mathrm{NLoS}}$ denotes the non-line-of-sight (NLoS) exponent. In addition, the noise power are set to $\left(\sigma^{\mathrm{ul}}\right)^2=\left(\sigma^{\mathrm{dl}}\right)^2=-10 \mathrm{dBm}$. All simulation results in the following are obtained by averaging over 100 experiments.
	\begin{figure}
		\centering
		\subfigure[MSE of AirComp versus the number of elements at STAR-RIS when $N=8$ and $K_m=4$.]{
			\centering
			\includegraphics[width=0.225\textwidth]{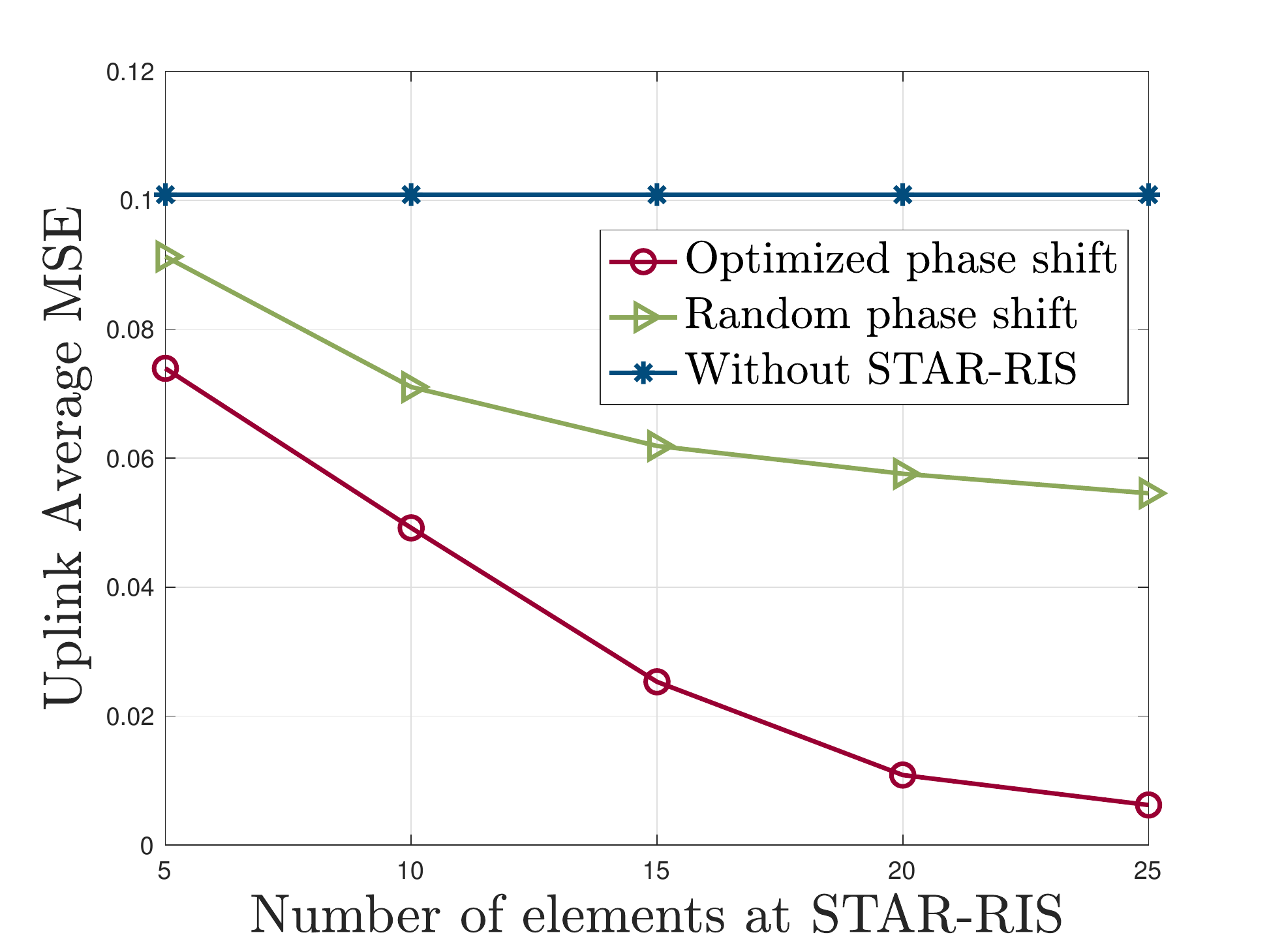}
		}
		\subfigure[Downlink MSE versus the number of elements at STAR-RIS when $N=8$ and $K_m=4$.]{
			\centering
			\includegraphics[width=0.225\textwidth]{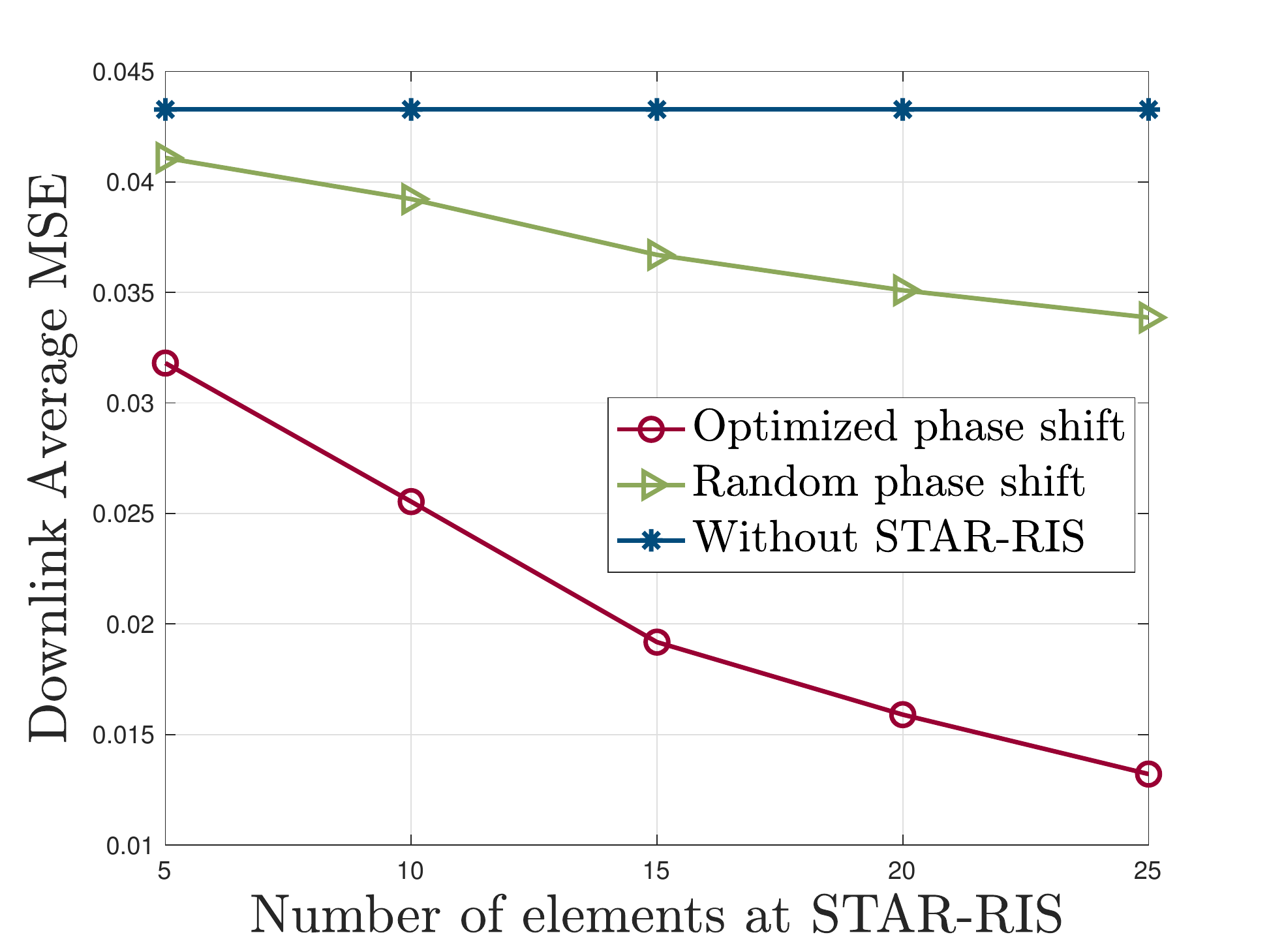}
		}
		\caption{Performance of uplink aggregation via AirComp under different settings.}
		\label{fig:mse}
	\end{figure}
	
	We first evaluate the performance of uplink aggregation using AirComp and downlink dissemination error by considering the MSE as the metric. As shown in Fig. \ref{fig:mse}, the MSE decreases as the number of STAR-RIS elements increases for both downlink and uplink transmission, indicating that STAR-RIS can effectively enhance the signal transmission quality, particularly when it has a large number of elements.
	
	We further evaluate the performance of our proposed STAR-RIS assisted vertical FL system, where $K_m = 4$ devices in each cell cooperatively train a regularized logistic regression model. The number of antennas at each BS is $N=8$, and the number of elements at STAR-RIS is $Q=10$. We simulate the image classification task on Fashion-MNIST dataset \cite{xiao2017fashion}. And we assume that each cell perform a different binary classification task for simplify (0-1 in cell 1, 2-3 in cell 2). The traditional binary cross-entropy loss function is given as
	\begin{equation}
		F(\bm{w}) = -\frac{1}{L} \sum_{i=1}^L \left[y^i\left(\bm{wx}^i\right) - \ln\left(1 + \exp(\bm{wx}^i)\right)\right]. \notag
	\end{equation}
	The learning rate $\mu^{(t)}$ is set to $0.01$.
	\begin{figure}
		\centering
		\subfigure[Training loss vs. Round]{
			\centering
			\includegraphics[width=0.225\textwidth]{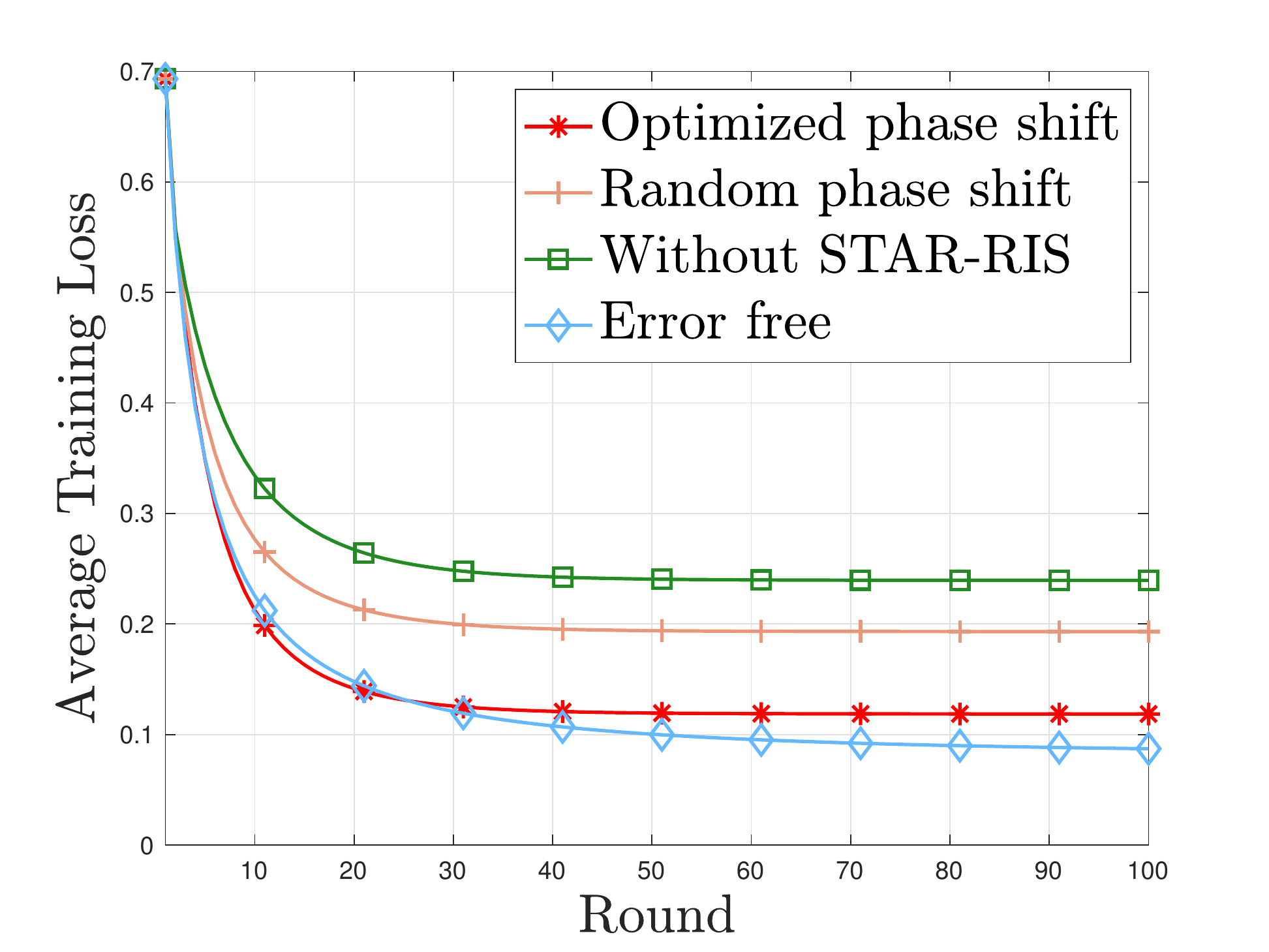}
		}
		\subfigure[Testing accuracy vs. Round]{
			\centering
			\includegraphics[width=0.225\textwidth]{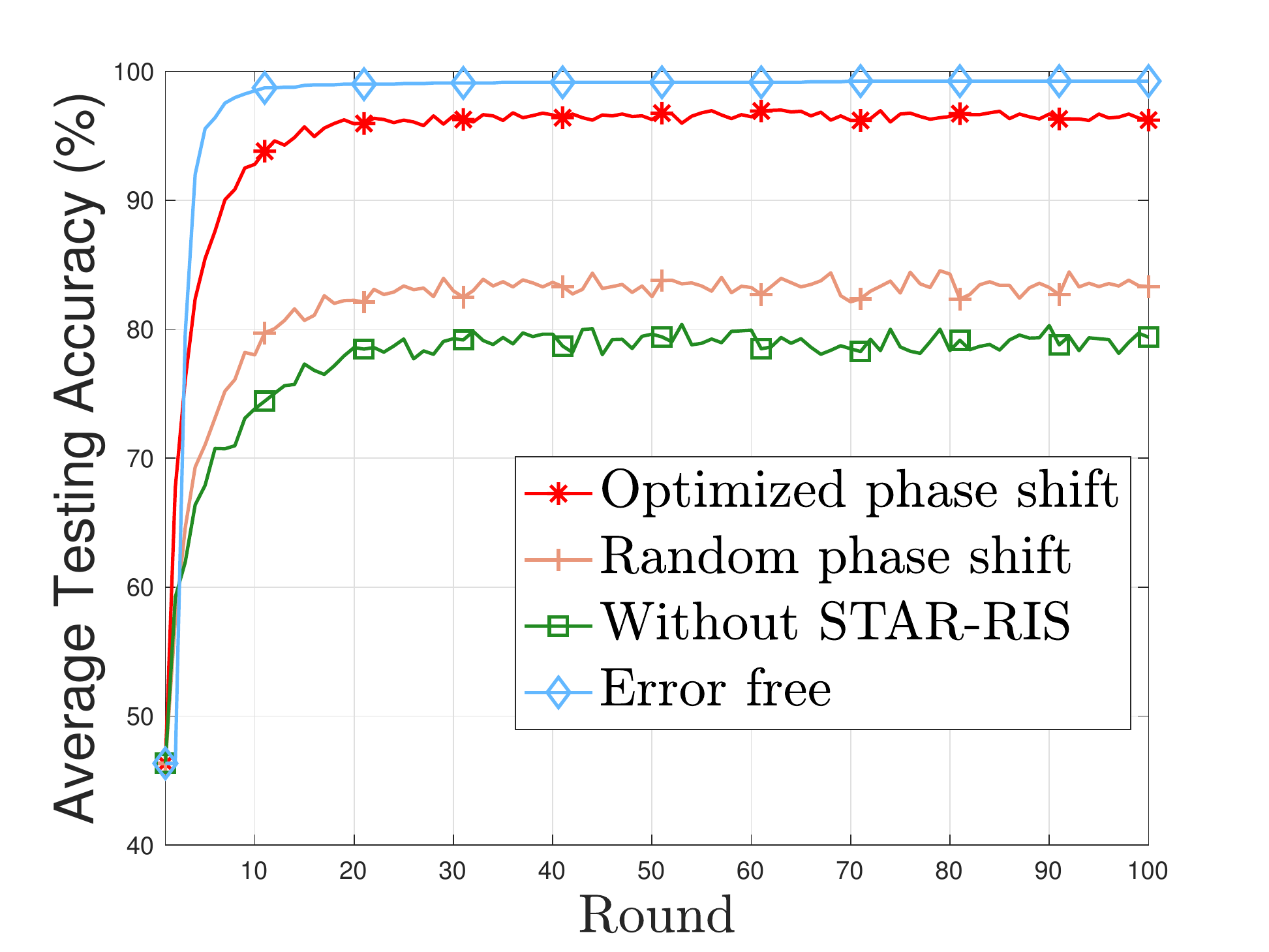}
		}
		\caption{Performance of AirComp assisted Vertical FL.}
		\label{fig:fl}
	\end{figure}
	
	We consider the noiseless case as the performance upper bound. Fig. \ref{fig:fl} shows that our proposed STAR-RIS assisted system converges quickly and achieves $96\%$ testing accuracy in inference, which is far ahead compared with the other two cases. And it is even close to the performance upper bound. 
	
	\section{Conclusion}
	In this paper, we proposed a STAR-RIS assisted AirComp-based vertical FL system in multi-cell networks. To be specific, a STAR-RIS is deployed at the cell edge to facilitate the completion of different FL tasks by each cell. The Pareto boundary of the gap region is introduced to characterize the trade-off of learning performance among cells. We then formulate an optimization problem to minimize the sum of error-induced gaps across all cells, which is then solved by SCA-based algorithms. Our simulation results demonstrate that the proposed STAR-RIS assisted system can significantly improve the learning performance in both training and inference phases thanks to its powerful capability of reducing the transmission errors. 
	
	\bibliographystyle{ieeetr}
	\bibliography{ref}

\begin{thebibliography}{10}

\bibitem{9606720}
K.~B. Letaief, Y.~Shi, J.~Lu, and J.~Lu, ``Edge artificial intelligence for 6g:
  Vision, enabling technologies, and applications,'' {\em IEEE J. Sel. Areas
  Commun.}, vol.~40, no.~1, pp.~5--36, 2022.

\bibitem{8808168}
K.~B. Letaief, W.~Chen, Y.~Shi, J.~Zhang, and Y.-J.~A. Zhang, ``The roadmap to
  6g: Ai empowered wireless networks,'' {\em IEEE Commun. Mag.}, vol.~57,
  no.~8, pp.~84--90, 2019.

\bibitem{shi2020communication}
Y.~Shi, K.~Yang, T.~Jiang, J.~Zhang, and K.~B. Letaief,
  ``Communication-efficient edge ai: Algorithms and systems,'' {\em IEEE
  Commun. Surveys Tuts.}, vol.~22, no.~4, pp.~2167--2191, 2020.

\bibitem{wang2020wireless}
Z.~Wang, Y.~Shi, Y.~Zhou, H.~Zhou, and N.~Zhang, ``Wireless-powered
  over-the-air computation in intelligent reflecting surface-aided iot
  networks,'' {\em IEEE Internet Things J.}, vol.~8, no.~3, pp.~1585--1598,
  2020.

\bibitem{yang2020federated}
K.~Yang, T.~Jiang, Y.~Shi, and Z.~Ding, ``Federated learning over-the-air
  computation,'' {\em IEEE Trans. Wireless Commun.}, vol.~19, no.~3,
  pp.~2022--2035, 2020.

\bibitem{yang2022differentially}
Y.~Yang, Y.~Zhou, Y.~Wu, and Y.~Shi, ``Differentially private federated
  learning via reconfigurable intelligent surface,'' {\em arXiv preprint
  arXiv:2203.17028}, 2022.

\bibitem{zhu2019broadband}
G.~Zhu, Y.~Wang, and K.~Huang, ``Broadband analog aggregation for low-latency
  federated edge learning,'' {\em IEEE Trans. Wireless Commun.}, vol.~19,
  no.~1, pp.~491--506, 2019.

\bibitem{xu2021bandwidth}
J.~Xu, H.~Wang, and L.~Chen, ``Bandwidth allocation for multiple federated
  learning services in wireless edge networks,'' {\em IEEE Trans. Wireless
  Commun.}, vol.~21, no.~4, pp.~2534--2546, 2021.

\bibitem{luo2021reconfigurable}
C.~Luo, X.~Li, S.~Jin, and Y.~Chen, ``Reconfigurable intelligent
  surface-assisted multi-cell {MISO} communication systems exploiting
  statistical {CSI},'' {\em IEEE Wireless Commun. Lett.}, vol.~10, no.~10,
  pp.~2313--2317, 2021.

\bibitem{huang2020holographic}
C.~Huang, S.~Hu, G.~C. Alexandropoulos, A.~Zappone, C.~Yuen, R.~Zhang,
  M.~Di~Renzo, and M.~Debbah, ``Holographic {MIMO} surfaces for 6{G} wireless
  networks: Opportunities, challenges, and trends,'' {\em IEEE Wireless
  Commun.}, vol.~27, no.~5, pp.~118--125, 2020.

\bibitem{liu2021star}
Y.~Liu, X.~Mu, J.~Xu, R.~Schober, Y.~Hao, H.~V. Poor, and L.~Hanzo, ``{STAR}:
  Simultaneous transmission and reflection for 360° coverage by intelligent
  surfaces,'' {\em IEEE Wireless Commun.}, vol.~28, no.~6, pp.~102--109, 2021.

\bibitem{9814484}
X.~Zeng, S.~Xia, K.~Yang, Y.~Wu, and Y.~Shi, ``Over-the-air computation for
  vertical federated learning,'' in {\em 2022 IEEE Int. Conf. Commun. Workshops
  (ICC Workshops)}, pp.~788--793, 2022.

\bibitem{liu2021reconfigurable}
H.~Liu, X.~Yuan, and Y.-J.~A. Zhang, ``Reconfigurable intelligent surface
  enabled federated learning: A unified communication-learning design
  approach,'' {\em IEEE Trans. Wireless Commun.}, vol.~20, no.~11,
  pp.~7595--7609, 2021.

\bibitem{wang2021federated}
Z.~Wang, J.~Qiu, Y.~Zhou, Y.~Shi, L.~Fu, W.~Chen, and K.~B. Letaief,
  ``Federated learning via intelligent reflecting surface,'' {\em IEEE Tran.
  Wireless Commun.}, vol.~21, no.~2, pp.~808--822, 2021.

\bibitem{li2019convergence}
X.~Li, K.~Huang, W.~Yang, S.~Wang, and Z.~Zhang, ``On the convergence of fedavg
  on non-iid data,'' {\em arXiv preprint arXiv:1907.02189}, 2019.

\bibitem{jorswieck2008complete}
E.~A. Jorswieck, E.~G. Larsson, and D.~Danev, ``Complete characterization of
  the pareto boundary for the {MISO} interference channel,'' {\em IEEE Trans.
  Signal Process.}, vol.~56, no.~10, pp.~5292--5296, 2008.

\bibitem{cao2020cooperative}
X.~Cao, G.~Zhu, J.~Xu, and K.~Huang, ``Cooperative interference management for
  over-the-air computation networks,'' {\em IEEE Trans. Wireless Commun.},
  vol.~20, no.~4, pp.~2634--2651, 2020.

\bibitem{xiao2017fashion}
H.~Xiao, K.~Rasul, and R.~Vollgraf, ``Fashion-mnist: a novel image dataset for
  benchmarking machine learning algorithms,'' {\em arXiv preprint
  arXiv:1708.07747}, 2017.

\end{thebibliography}
\end{document}